# Four five-parametric and five four-parametric independent confluent Heun potentials for the stationary Klein-Gordon equation

**A.S. Tarloyan[1], T.A. Ishkhanyan[1,2], and A.M. Ishkhanyan[1]**

[1]Institute for Physical Research, NAS of Armenia, 0203 Ashtarak, Armenia
[2]Moscow Institute of Physics and Technology, Dolgoprudny, Moscow Region, 141700 Russia

We present in total fifteen potentials for which the stationary Klein-Gordon equation is solvable in terms of the confluent Heun functions. Because of the symmetry of the confluent Heun equation with respect to the transposition of its regular singularities, only nine of the potentials are independent. Four of these independent potentials are five-parametric. One of them possesses a four-parametric ordinary hypergeometric sub-potential, another one possesses a four-parametric confluent hypergeometric sub-potential, and one potential possesses four-parametric sub-potentials of both hypergeometric types. The fourth five-parametric potential has a three-parametric confluent hypergeometric sub-potential, which is, however, only conditionally integrable. The remaining five independent Heun potentials are four-parametric and have solutions only in terms of irreducible confluent Heun functions.

## 1. Introduction

The Klein-Gordon equation [1,2] is a relativistic version of the Schrödinger equation that describes the behavior of spinless particles. The equation has a large range of applications in contemporary physics, including particle physics, astrophysics, cosmology, classical mechanics, etc. (see [1-4] and references therein). For the stationary problems, when the Hamiltonian does not depend on time, particular solutions can be obtained by applying the separation of variables that reduces the problem to the solution of the stationary Klein-Gordon equation. This approach is widely used to treat particles in various external fields or curved space-time using functions of the hypergeometric [5-14] or the Heun [15-20] classes.

In the present paper we consider the reduction of the one-dimensional stationary Klein-Gordon equation to the single-confluent Heun equation [21-22]. This equation possesses two regular singular points located at finite points of the complex $z$-plane and an irregular singularity of $s$-rank 2 at infinity [22]. Owing to such a structure of the singularities, the confluent Heun equation directly incorporates, by simple choices of the involved exponent parameters, the Gauss ordinary hypergeometric and the Kummer confluent hypergeometric equations as well as the algebraic form of the Mathieu equation and several other familiar equations. Because of the richer structure of the singularities, it is clear that the confluent Heun equation is potent to suggest a set of potentials that cannot be treated by the hypergeometric equations in reasonable limits. In the meantime, since the parameters standing for different singularities are clearly separated so that the influence of the each feature originating from a particular singularity is well identified, it is expected that the confluent Heun generalizations will suggest a clear route to follow the details relevant to a particular prototype hypergeometric or Mathieu potential.

We show that, to derive energy-independent potentials that are in addition proportional to an energy-independent continuous parameter and for which the potential shape is independent of the latter parameter, there exist only 15 permissible choices for the coordinate transformation. Each of these transformations leads to a four- or five-parametric potential solvable in terms of the confluent Heun functions. However, because of the symmetry of the confluent Heun equation with respect to the transposition of its regular singularities, only nine of these potentials are independent. Four of the independent potentials are five-parametric and five others are four-parametric.

The five-parametric Heun potentials all possess hypergeometric sub-potentials while the four-parametric ones do not. One of the five-parametric potentials has a four-parametric sub-potential solvable in terms of the Gauss hypergeometric function, another potential has a four-parametric sub-potential solvable in terms of the Kummer confluent hypergeometric function and there is a potential that possesses four-parametric sub-potentials of both hypergeometric types. Finally, the fourth five-parametric Heun potential possesses a three-parametric confluent hypergeometric sub-potential which is, however, only conditionally integrable in the sense that in this case the potential cannot be presented as being proportional to a parameter and having a shape that is independent of that parameter.

Methodologically, we follow the route for reduction of the problem to a target second-order ordinary differential equation having rational coefficients applied for the case of the Schrödinger equation in [23-26] and to the quantum two-state problem in [27-29]. The basic approach suggested in [23] for construction of exactly integrable energy-independent potentials by transforming the dependent and independent variables rests on the observation that, if a potential is proportional to an energy-independent parameter and the potential shape is independent of both energy and that parameter, then the logarithmic derivative $\rho'(z)/\rho(z)$ of the function $\rho = z'(x)$, where $z = z(x)$ is the coordinate transformation, cannot have poles other than the finite singularities of the target equation. It then follows that the function $\rho$ should necessarily be of the Manning form $\rho(z) = \Pi_i (z-z_i)^{A_i}$ [30] with $z_i$ being the finite singularities of the target equation and the exponents $A_i$ all being integers or half-integers.

## 2. Confluent Heun potentials

The one-dimensional Klein-Gordon equation for a particle of rest mass $m$ and energy $E$ in a scalar potential field $V(x)$ is written as [1]

$$\frac{d^2\psi}{dx^2} + \frac{1}{\hbar^2 c^2}\left((E-V(x))^2 - m^2 c^4\right)\psi = 0. \tag{1}$$

Applying the independent variable transformation $z = z(x)$, this equation is rewritten for the new argument $z$ as

$$\psi_{zz} + \frac{\rho_z}{\rho}\psi_z + \frac{1}{\hbar^2 c^2}\frac{(E-V(z))^2 - m^2 c^4}{\rho^2}\psi = 0, \tag{2}$$

where $\rho = dz/dx$ and the lowercase Latin index denotes differentiation. Further transformation of the dependent variable $\psi = \varphi(z) u(z)$ reduces this equation to the following one for the new dependent variable $u(z)$:

$$u_{zz} + \left(2\frac{\varphi_z}{\varphi} + \frac{\rho_z}{\rho}\right) u_z + \left(\frac{\varphi_{zz}}{\varphi} + \frac{\rho_z}{\rho}\frac{\varphi_z}{\varphi} + \frac{1}{\hbar^2 c^2}\frac{(E-V(z))^2 - m^2 c^4}{\rho^2}\right) u = 0. \tag{3}$$

This equation becomes the single-confluent Heun equation [21-22]:

$$u_{zz} + \left(\frac{\gamma}{z} + \frac{\delta}{z-1} + \varepsilon\right) u_z + \frac{\alpha z - q}{z(z-1)} u = 0, \tag{4}$$

if

$$\frac{\gamma}{z} + \frac{\delta}{z-1} + \varepsilon = 2\frac{\varphi_z}{\varphi} + \frac{\rho_z}{\rho} \tag{5}$$

and

$$\frac{\alpha z - q}{z(z-1)} = \frac{\varphi_{zz}}{\varphi} + \frac{\rho_z}{\rho}\frac{\varphi_z}{\varphi} + \frac{1}{\hbar^2 c^2}\frac{(E-V(x))^2 - m^2 c^4}{\rho^2}. \tag{6}$$

Resolving equation (5) for $\varphi$:

$$\varphi = \frac{e^{\varepsilon z/2}}{\sqrt{\rho(z)}} z^{\frac{\gamma}{2}} (z-1)^{\frac{\delta}{2}}, \tag{7}$$

and substituting this into equation (6), we get

$$\frac{C_0 + C_1 z + C_2 z^2 + C_3 z^3 + C_4 z^4}{z^2 (z-1)^2} + \left(\frac{\rho_z^2}{4\rho^2} - \frac{\rho_{zz}}{2\rho}\right) + \frac{1}{\hbar^2 c^2}\frac{(E-V(z))^2 - m^2 c^4}{\rho^2} = 0, \tag{8}$$

where the constants $C_{0,1,2,3,4}$ are defined by the parameters of the confluent Heun equation.

We suppose that the potential is energy-independent and is proportional to an independent parameter $\mu$: $V(z) = \mu f(z)$, with a potential shape $f(z)$ that is independent of that parameter: $f \neq f(\mu)$. A key observation of [23] for this case is that, for an $E$-independent coordinate transformation $z = z(x)$, equation (8) can be satisfied only if

$$z'(x) = \rho = z^{m_1}(z-1)^{m_2}/\sigma \qquad (9)$$

with integer or half-integer $m_{1,2}$.

The lines leading to this conclusion are as follows. Taking the second derivative of equation (8) with respect to $E$, we see that

$$\frac{1}{\rho^2} = \frac{r_0 + r_1 z + r_2 z^2 + r_3 z^3 + r_4 z^4}{z^2(z-1)^2} \equiv \frac{r(z)}{z^2(z-1)^2}, \qquad (10)$$

where $r_i = (c^2\hbar^2/2)d^2C_i/dE^2$. Rewriting now the polynomial $r(z)$ as $r(z) = \prod_i (r - s_i)$, taking the limit $E, \mu \to 0$ and applying the identity

$$\frac{\rho_{zz}}{\rho} = \left(\frac{\rho_z}{\rho}\right)_z + \left(\frac{\rho_z}{\rho}\right)^2, \qquad (11)$$

we see that the roots $s_i$ are 0 or 1 (in other words, one can say that the logarithmic derivative $\rho_z/\rho$ cannot have poles other than the finite singularities $z = 0,1$ of the confluent Heun equation). With this and equation (10), we arrive at equation (9) with integer or half-integer $m_{1,2}$ and arbitrary $\sigma$. Besides, since $z^2(z-1)^2/\rho^2 = z^{2-2m_1}(z-1)^{2-2m_2}$ is a polynomial of at most fourth degree, we have the inequalities $-1 \leq m_{1,2} \leq 1$ and $0 \leq m_1 + m_2 \leq 2$. This leads to 15 possible sets of $m_{1,2}$ shown in Fig.1. We note that, because of symmetry of the confluent Heun equation with respect to the transposition $z \leftrightarrow 1-z$, only nine of these cases are independent. The independent cases are marked in the figure by filled shapes.

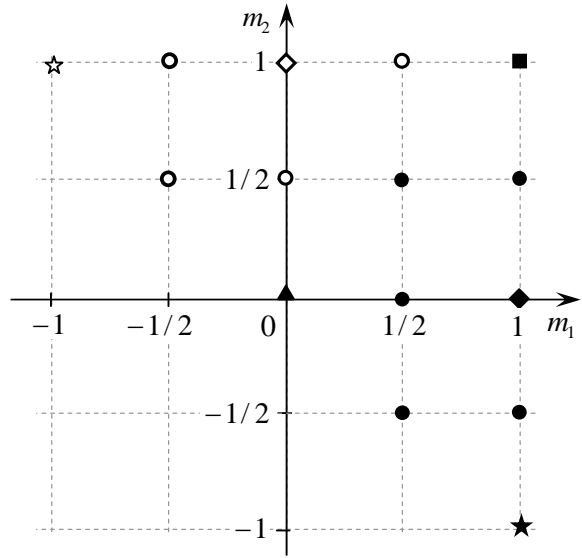

Fig. 1. Fifteen possible pairs $m_1, m_2$. The nine independent cases are marked by filled shapes. The cases possessing ordinary or confluent hypergeometric sub-potentials are marked by square or triangle, respectively. The rhombs indicate the cases that possess hypergeometric sub-potentials of both types. The asterisks mark the conditionally integrable cases.

The next step is matching the cross-term $-2EV(z)/\rho^2$ and the term $V^2(z)/\rho^2$ with the rest in equation (8). Taking the first derivative with respect to $E$, we first get

$$z^2(z-1)^2 \frac{V(z)}{\rho^2} = v_0 + v_1 z + v_2 z^2 + v_3 z^3 + v_4 z^4 \equiv v(z), \qquad (12)$$

and further taking the limit $E \to 0$ find that the parameters $v_{0,1,2,3,4}$ should be so chosen that

$$z^2(z-1)^2 \frac{V^2(z)}{\rho^2} = z^{2m_1-2}(z-1)^{2m_2-2} v^2(z) \equiv w(z) \qquad (13)$$

is a polynomial of at most fourth degree. By direct inspection it is then shown that the last requirement is fulfilled only for certain permissible sub-sets of the parameters $v_{0,1,2,3,4}$. For the nine independent cases of $m_{1,2}$ the resultant potentials can be conveniently written in the form presented in Table 1. Four of the independent potentials are five-parametric, while the remaining five potentials are four-parametric ($V_{0,1,2}, x_0, \sigma$ are arbitrary complex constants). The types of the hypergeometric sub-potentials for the four five-parametric cases possessing such sub-potentials are indicated in the last column of the table.

| N | $m_1, m_2$ | Potential $V(z)$ | Coordinate transformation $z(x)$ or $x(z)$ | Hypergeom. sub-potential |
|---|---|---|---|---|
| 1 | 0, 0 | $V_0 + \frac{V_1}{z} + \frac{V_2}{z-1}$ | $z(x) = \frac{x-x_0}{\sigma}$ | $_1F_1$ [31] |
| 2 | $1/2, -1/2$ | $V_0 + \frac{V_1}{z-1}$ | $x(z) = x_0 + \sigma\left(\sqrt{z(z-1)} - \sinh^{-1}\left(\sqrt{z-1}\right)\right)$ | --- |
| 3 | $1/2, 0$ | $V_0 + \frac{V_1}{z-1}$ | $z(x) = \frac{(x-x_0)^2}{4\sigma^2}$ | --- |
| 4 | $1/2, 1/2$ | $V_0 + V_1 z$ | $z(x) = \cosh^2\left(\frac{x-x_0}{2\sigma}\right)$ | --- |
| 5 | $1, -1$ | $V_0 + \frac{V_1}{z-1} + \frac{V_2}{(z-1)^2}$ | $x(z) = x_0 + \sigma(z - \log(z))$<br>Lambert W: $z(x) = -W\left(-e^{-(x-x_0)/\sigma}\right)$ | Conditionally solvable: $_1F_1$ |
| 6 | $1, -1/2$ | $V_0 + \frac{V_1}{z-1}$ | $x(z) = x_0 + 2\sigma\left(\sqrt{z-1} - \tan^{-1}\left(\sqrt{z-1}\right)\right)$ | --- |
| 7 | $1, 0$ | $V_0 + V_1 z + \frac{V_2}{z-1}$ | $z(x) = e^{\frac{x-x_0}{\sigma}}$ | $_1F_1$ [32]<br>$_2F_1$ [33] |
| 8 | $1, 1/2$ | $V_0 + V_1 z$ | $z(x) = \sec^2\left(\frac{x-x_0}{2\sigma}\right)$ | --- |
| 9 | $1, 1$ | $V_0 + V_1 z + V_2 z^2$ | $z(x) = \frac{1}{e^{(x-x_0)/\sigma} + 1}$ | $_2F_1$ [35] |

Table 1. Nine independent potentials together with the coordinate transformation. $V_{0,1,2}$ and $x_0, \sigma$ are arbitrary (complex) constants.

The solution of the problem is readily written taking the pre-factor $\varphi(z)$ as

$$\varphi = e^{\alpha_0 z} z^{\alpha_1} (z-1)^{\alpha_2}. \qquad (14)$$

Then, collecting the coefficients at powers of $z$ in the numerators of equations (5) and (6), we get eight equations which are linear for the five parameters $\gamma, \delta, \varepsilon, \alpha, q$ of the confluent Heun function $u(z)$ and are quadratic for the three parameters $\alpha_{0,1,2}$ of the pre-factor. Resolving these equations, we finally get that the solution of the stationary Klein-Gordon equation (1) is explicitly written in terms of the confluent Heun function as

$$\psi = e^{\alpha_0 z} z^{\alpha_1} (z-1)^{\alpha_2} H_C(\gamma, \delta, \varepsilon; \alpha, q; z) \qquad (15)$$

with the involved parameters being given by the equations

$$\gamma = 2\alpha_1 + m_1, \quad \delta = 2\alpha_2 + m_2, \quad \varepsilon = 2\alpha_0, \qquad (16)$$

$$\alpha = \alpha_0 \left(m_1 + m_2 + 2(\alpha_1 + \alpha_2 - \alpha_0)\right) + \frac{1}{\hbar^2 c^2}\left(\left(E^2 - m^2 c^4\right) r_3 - 2E v_3 + w_3\right), \qquad (17)$$

$$q = \alpha_1 (2 - m_1 - m_2) + (2\alpha_1 + m_1)(\alpha_0 - \alpha_1 - \alpha_2) + \frac{1}{\hbar^2 c^2}\left(\left(E^2 - m^2 c^4\right) r_1 - 2E v_1 + w_1\right). \qquad (18)$$

The equations for the exponents $\alpha_{0,1,2}$ read

$$\alpha_0^2 + \frac{1}{\hbar^2 c^2}\left(\left(E^2 - m^2 c^4\right) r_4 - 2E v_4 + w_4\right) = 0, \qquad (19)$$

$$\alpha_1^2 - \alpha_1 (1 - m_1) + \frac{1}{\hbar^2 c^2}\left(\left(E^2 - m^2 c^4\right) r(0) - 2E v(0) + w(0)\right) = 0, \qquad (20)$$

$$\alpha_2^2 - \alpha_2 (1 - m_2) + \frac{1}{\hbar^2 c^2}\left(\left(E^2 - m^2 c^4\right) r(1) - 2E v(1) + w(1)\right) = 0 \qquad (21)$$

and the auxiliary parameters $r_{0,1,2,3,4}$, $v_{0,1,2,3,4}$ and $w_{0,1,2,3,4}$ for each row of Table 1 are readily calculated through the definitions (10), (12) and (13):

$$r(z) = r_0 + r_1 z + r_2 z^2 + r_3 z^3 + r_4 z^4 = z^{2-2m_1}(z-1)^{2-2m_2}, \qquad (22)$$

$$v_0 + v_1 z + v_2 z^2 + v_3 z^3 + v_4 z^4 = r(z) V(z), \qquad (23)$$

$$w_0 + w_1 z + w_2 z^2 + w_3 z^3 + w_4 z^4 = r(z) V^2(z) \qquad (24)$$

The coordinate transformation $x(z)$ or $z(x)$ calculated using equation (9) is presented in the third column of Table 1.

The derived solution applies to any set of the involved parameters. It should be stressed that the parameters in general may be chosen complex. For example, putting $x_0 \to x_0 + i\pi\sigma$, one may change the sign of the exponents involved in the coordinate transformations in the 7th and 9th rows of Table 1.

## 3. Hypergeometric sub-potentials

Consider the hypergeometric reductions of the above confluent Heun potentials. We first demand for the parameters of the hypergeometric sub-potentials to be independent of each other. Then, the results are as follows.

The confluent Heun equation is reduced to the Kummer confluent hypergeometric equation if $\delta = 0 \cup q = \alpha$ or $\gamma = 0 \cup q = 0$. Examining these two possibilities through equations (16)-(24), one readily reveals that this is possible only in three cases when $m_{1,2}$ are integers (half-integer $m_{1,2}$ lead to constant potentials) obeying the inequality $0 \leq m_1 + m_2 \leq 1$. Because of the symmetry of the potentials with respect to the transposition $m_1 \leftrightarrow m_2 \cup z \leftrightarrow 1-z$, the number of the independent cases is reduced to two. These are the Coulomb potential [31] and the exponential potential shown in Table 2, where the numbers in the first column indicate the number of the confluent Heun potential to which the particular hypergeometric sub-potential belongs. The first of the two confluent hypergeometric potentials, the Coulomb potential, has been applied in the past by many authors starting from the early days of quantum mechanics [1,2]. The second potential can be viewed as a truncated one-term version of the Morse potential [32].

The confluent Heun equation is reduced to the Gauss ordinary hypergeometric equation if $\varepsilon = \alpha = 0$. It then follows from Eqs. (16), (17), (19) that in this case $r_3 = v_3 = w_3 = 0$ and $r_4 = v_4 = w_4 = 0$, so that the polynomials $r(z)$, $v(z)$ and $w(z)$ are of the second degree. Accordingly, $m_{1,2}$ obey the inequality $1 \leq m_1 + m_2 \leq 2$ (see Eq. (22)). Hence, hypergeometric sub-potentials may exist only for the six sets $m_{1,2}$ close to the upper right-hand corner in Fig.1. Because of the symmetry of the hypergeometric equation with respect to the transposition $z \leftrightarrow 1-z$, the number of the independent cases is reduced to four. A closer inspection further reveals that the sets with $m_1 = 1/2$ or $m_2 = 1/2$ do not lead to non-constant potentials. Thus, we arrive at two four-parametric potentials presented in Table 2. The first one is identified as a version of the Hulthén potential [33], which presents a one-term four-parametric specification of the two-term five-parametric Eckart potential [34]. The second potential is the Woods-Saxon potential [35], which is again a four-parametric one-term specification of the Eckart potential. We note that the two hypergeometric sub-potentials are transformed into each other by simple change $x_0 \to x_0 + i\pi\sigma$. Thus, there exists only one independent ordinary hypergeometric potential. This potential has been explored in the past by many authors on several occasions (see, e.g., [6-9]).

| N | $m_1, m_2$ | Potential $V(z)$ | Coordinate transformation $z(x)$ | Reference |
|---|---|---|---|---|
| 1 | 0, 0 | $V_0 + \dfrac{V_1}{z}$ | $z(x) = \dfrac{x-x_0}{\sigma}$ | $_1F_1$ Coulomb [31] |
| 7 | 1, 0 | $V_0 + V_1 z$ | $z(x) = e^{\frac{x-x_0}{\sigma}}$ | $_1F_1$ Exponential (Morse [32]) |
| 7 | 1, 0 | $V_0 + \dfrac{V_1}{z-1}$ | $z(x) = e^{\frac{x-x_0}{\sigma}}$ | $_2F_1$ Hulthén [33] (Eckart [34]) |
| 9 | 1, 1 | $V_0 + V_1 z$ | $z(x) = \dfrac{1}{e^{(x-x_0)/\sigma}+1}$ | $_2F_1$ Woods-Saxon [35] (Eckart [34]) |

Table 2. Confluent and ordinary hypergeometric potentials. The two ordinary hypergeometric potentials are transformed into each other by the change $x_0 \to x_0 + i\pi\sigma$.

We would like to conclude this section by noting that if a weaker requirement of *conditional* solvability (that is if a parameter of the potential is fixed to a specific value or if the parameters standing for characteristics of different physical origin are dependent) is examined, there may exist other hypergeometric sub-potentials. An example of this kind of sub-potentials is as follows. Consider the case $m_{1,2} = (1,-1)$. It is then readily verified that for

$$V_1 = -\frac{c\hbar}{\sqrt{3}\sigma}, \quad V_2 = -\frac{\sqrt{3}c\hbar}{2\sigma}, \tag{25}$$

that is for the potential

$$V = V_0 - \frac{c\hbar}{\sqrt{3}\sigma}\left(\frac{1}{z-1} + \frac{3/2}{(z-1)^2}\right), \quad z = -W\left(-e^{-\frac{x-x0}{\sigma}}\right) \tag{26}$$

it holds $\delta = 0$ and $\alpha = q$ so that the confluent Heun equation (4) is reduced to the scaled Kummer confluent hypergeometric equation. This is a conditionally integrable potential since the interaction strengths $V_{1,2}$ depend on the space scale $\sigma$. More strictly, this potential cannot be presented as $V = \mu f(z)$ with $f \neq f(\mu)$. This three-parametric potential ($V_0, x_0, \sigma$ are arbitrary) and its counterpart for $m_{1,2} = (-1,1)$ are marked in Fig.1 by asterisks.

Choosing $x_0 = -\sigma$ and

$$V_0 = \frac{c\hbar}{2\sqrt{3}\sigma}, \tag{27}$$

we get a single-parametric potential defined for a positive $\sigma$ on the positive half-axis $x > 0$ that has a singularity at the origin and vanishes at infinity (Fig.2):

$$V = V_0 \frac{z(z-4)}{(z-1)^2}, \quad z = -W\left(-e^{-1-x/\sigma}\right). \tag{28}$$

In the vicinity of the origin the behavior of the potential is Coulomb-like:

$$V|_{x\to 0} = -\frac{\sqrt{3}c\hbar/4}{x} + O(1), \qquad (29)$$

while at infinity the potential vanishes exponentially:

$$V|_{x\to +\infty} = -\frac{2c\hbar}{\sqrt{3}\sigma}e^{-x/\sigma} + O(e^{-2x/\sigma}). \qquad (30)$$

The solution of the Klein-Gordon equation for this potential is explicitly written as

$$\psi = z^{\alpha_1}(1-z)^{1/2}e^{\varepsilon z/2}\,_1F_1(a;1+2\alpha_1;-\varepsilon z) \qquad (31)$$

with

$$\alpha_1 = \pm\frac{\sigma}{c\hbar}\sqrt{m^2c^4 - E^2}, \quad \varepsilon = \pm\frac{2\sigma}{c\hbar}\sqrt{m^2c^4 - (E-V_0)^2} \qquad (32)$$

and

$$a = \alpha_1 + \left(\frac{1-\varepsilon}{2} - \frac{2}{3\varepsilon}\right) + \frac{m^2c^4 - E^2}{3\varepsilon V_0^2} + \frac{E}{\varepsilon V_0}. \qquad (33)$$

Here any combination of the signs + or − is applicable for $\alpha_1$ and $\varepsilon$. We note that by choosing different signs we get different independent fundamental solutions.

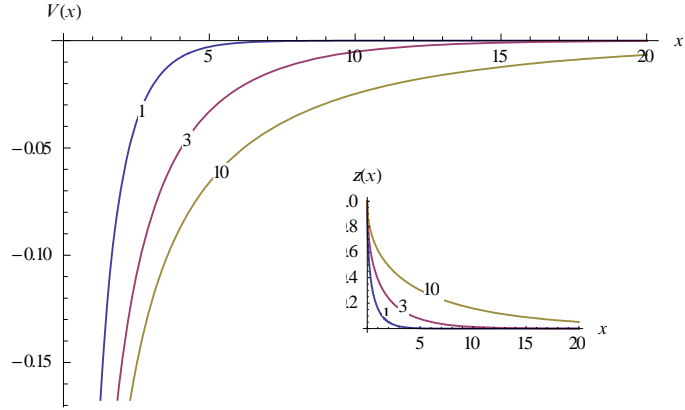

Fig.1. Potential (28) for $\sigma = 1, 3, 10$ ($c = \hbar = 1$). The inset presents the coordinate transformation $z(x)$.

**Discussion**

There have been many studies employing reductions of the stationary Klein-Gordon equation to the Heun class of differential equations (for the reductions to the single-confluent Heun equation see, for instance, [15-19] and for a list of sub-cases belonging to the hypergeometric class see [14]). However, to the best of our knowledge, there has not been a general discussion of this question.

In the present paper we have presented a systematic treatment using the general approach developed in [23-26] for the stationary Schrödinger equation and in [27-29] for the time-dependent quantum two-state problem (see also [36-38]). This is an approach for searching for exactly solvable energy-independent potentials via energy-independent coordinate transformation that is designed for identification of solvable potentials that are proportional to an energy-independent parameter and have a shape that does not depend on that parameter [23]. It should be noted, however, that the technique is also potent to generate conditionally integrable potentials - we have presented an example of this kind of potentials (see another example in [25], several other examples are presented in [29,36,37]).

Discussing the representative example of the single-confluent Heun equation which directly incorporates the two hypergeometric equations, we have shown that there exist in total fifteen permissible choices for the coordinate transformation each leading to a four- or five- parametric potential. Because of the symmetry of the single-confluent Heun equation with respect to the transposition $z \to 1-z$, only nine of these potentials are independent. Five of the independent potentials are four-parametric. A peculiarity of these potentials is that they do not posses hypergeometric sub-potentials. Four other independent confluent Heun potentials are five-parametric. These potentials present distinct generalizations of certain hypergeometric potentials.

Among the independent five-parametric confluent Heun potentials, one potential extends the Coulomb confluent hypergeometric potential [31], another one extends the Hulthén ordinary hypergeometric potential [33], and there is a potential that possesses four-parametric sub-potentials of both hypergeometric types (exponential potential, which is a truncated version of the Morse confluent hypergeometric potential [32], and the Woods-Saxon ordinary hypergeometric potential [35]). The fourth five-parametric Heun potential possesses a confluent hypergeometric sub-potential which is a conditionally integrable in the sense that it cannot be presented as being proportional to an energy-independent parameter and having a shape that is independent of that parameter. This is a three-parametric potential explicitly given through a coordinate transformation written in terms of the Lambert $W$-function [39-40]. The Schrödinger counterpart of this potential has been presented in [41].

We would like to conclude by noting that the applied approach and the lines of the presented analysis are rather general and can be extended to other target equations or to treat other structurally similar problems, for instance, to derive the Heun solutions of the Klein-Gordon equation on manifolds with variable geometry [42] or to identify the potentials for which the Klein-Gordon and other relativistic quantum mechanical wave equations [1,3] are solvable in terms of the general or multiply-confluent Heun functions.

**Acknowledgments**

This research has been conducted within the scope of the International Associated Laboratory (CNRS-France & SCS-Armenia) IRMAS. The work has been supported by the Armenian State Committee of Science (SCS Grants no. 13YR-1C0055 and no. 13RB-052).